\documentclass[twocolumn,superscriptaddress,10pt,aps,showpacs,reprint]{revtex4-1}

\usepackage{amsmath}
\usepackage{amssymb}
\usepackage{microtype}
\usepackage{graphicx}

\usepackage{color} 

\usepackage{calligra}
\DeclareMathAlphabet{\mathcalligra}{T1}{calligra}{m}{n}
\DeclareFontShape{T1}{calligra}{m}{n}{<->s*[2.2]callig15}{}
\newcommand{\scripty}[1]{\ensuremath{\mathcalligra{#1}}}

\begin{document}

\title{Emergent Structural Mechanisms for High-Density Collective Motion Inspired by Human Crowds}

\author{Arianna Bottinelli}
\email{arianna.bottinelli@math.uu.se}
\affiliation{Mathematics Department, Uppsala University, L\"agerhyddsv\"agen 1, Uppsala 75106, Sweden}

\author{David T. J. Sumpter}
\email{david.sumpter@math.uu.se}
\affiliation{Mathematics Department, Uppsala University, L\"agerhyddsv\"agen 1, Uppsala 75106, Sweden}

\author{Jesse L. Silverberg}
\email{jesse.silverberg@wyss.harvard.edu}
\affiliation{Wyss Institute for Biologically Inspired Engineering, Harvard University, Boston MA, USA.}

\begin{abstract}
Collective motion of large human crowds often depends on their density.  In extreme cases like heavy metal concerts and Black Friday sales events, motion is dominated by physical interactions instead of conventional social norms.  Here, we study an active matter model inspired by situations when large groups of people gather at a point of common interest.  Our analysis takes an approach developed for jammed granular media and identifies Goldstone modes, soft spots, and stochastic resonance as structurally-driven mechanisms for potentially dangerous emergent collective motion. 
\end{abstract}


\maketitle

Studies of collective motion cover a broad range of systems including humans, fish, birds, locusts, cells, vibrated rice, colloids, actin-myosin networks, and even robots \cite{camazine2003self, sumpter2010collective, vicsek2012collective}.  Often, theoretical models of these active matter systems take a Newtonian approach by calculating individual trajectories generated \textit{in silico} from the sum of forces acting on each of $N$ particles \cite{vicsek2012collective}.  For the work focusing on humans, social interactions such as collision avoidance, tendencies to stay near social in-group members, directional alignment, and preference for personal space have been examined to understand their role in emergent behavior \cite{helbing1995social, helbing2011pedestrian, xi2011integrated, moussaid2011simple}.  Generally, these studies show order-disorder transitions are driven by the competition between social interactions and randomizing forces \cite{Silverberg2013, Helbing2000PRL}.  Moreover, these models have been incorporated into predictive tools used to enhance crowd management strategies at major organized gatherings.  In extreme social situations such as riots, protests, and escape panic, however, the validity of this approach is diminished \cite{Helbing2007, duives2013state, Helbing2000}.  Conventional social interactions no longer apply to individual people \cite{fruin1993causes}, and the actual collective behavior can be quite different from model predictions \cite{zeitz2009crowd, heide2004common}.

\begin{figure*}
\includegraphics[width=1\textwidth]{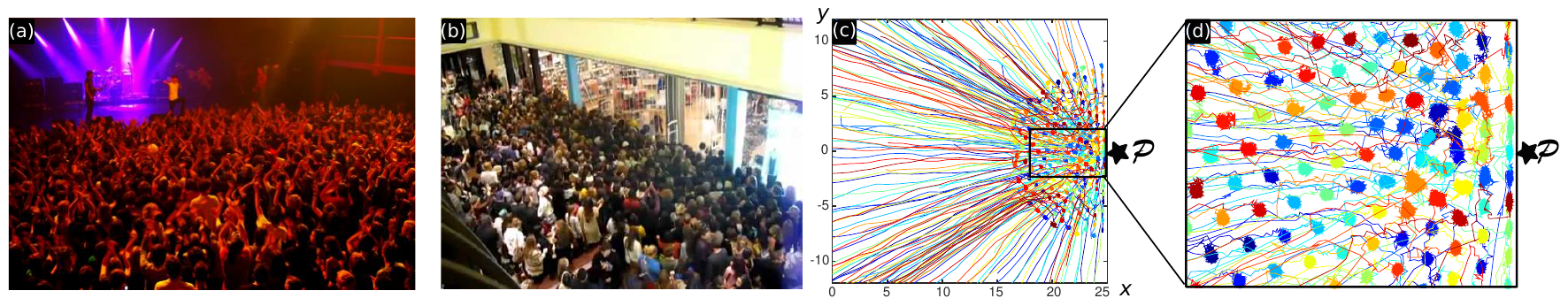}
\caption{(color online). Large dense groups of people give rise to emergent collective motion.  (a) Attendees gather near the stage at a heavy metal concert.  Credit: Ulrike Biets. (b) Customers gather for Black Friday sale to purchase low-cost consumer goods. Credit: Jerry Bailey.  (c) Simulated trajectories of SPPs aggregating near a point of interest ${\cal P}$ located at the right-most edge of a simulation box.  (d) Zoom in of trajectories show SPPs self-organize into a densely packed disordered aggregate.}
\label{fig:1}
\end{figure*}

Situations involving large groups of people packed at high-densities provide a unique view of the emergent collective behavior in extreme circumstances \cite{Helbing2007, Silverberg2013}.  For example, attendees at heavy metal concerts often try to get as close as possible to the stage, but are unable to do so due to the shear number of people trying to attain the same goal [Fig.~\ref{fig:1}(a)].  Consequently, the audience in this region of the concert venue becomes a densely packed shoulder-to-shoulder group with little room for individuals to freely move.  Often, the stresses involved become dangerously high and security professionals standing behind physical barriers are required to pull audience members from the crowd for medical attention \cite{janchar2000mosh}.  At Black Friday sales events, we find a similar situation when individuals seeking low-cost consumer goods congregate at the entrance of a store before it opens [Fig.~\ref{fig:1}(b)].  As documented in many news reports and online videos, these events can have tragic outcomes in the critical moments after the doors open and the crowd surges forward leading to increased risk of stampedes and trampling.

In extreme situations involving large high-density crowds, physical interaction between contacting bodies and the simultaneous collective desire of each individual to get to a stage, through a door, or to a particular location become the dominant considerations \cite{helbing2011pedestrian, Helbing2000, bohannon2005directing}. To generically capture these scenarios, we use a conventional force-based active matter model for human collective motion, but remove terms that account for social interaction.  With this simplification, we have an \textit{asocial} model for human collective behavior describing people aggregating around a common point of interest ${\cal P}$.  Here, we place ${\cal P}$ at the side of a 2D $L \times L$ simulation box [Fig.~\ref{fig:1}(c)].  In this framework, each person $i$ is modeled as a disk with radius $r_0 \ll L$ positioned at a point $\vec{r}_i(t)$ subject to pairwise soft-body respulsive collision forces $\vec{F}_i^{\rm repulsion}$, a self-propulsion force $\vec{F}_i^{\rm propulsion}$, random force fluctuations from environmental stimuli $\vec{F}_i^{\rm noise}$, and a rigid-wall collision force $\vec{F}_i^{\rm wall}$.  

For each of the $N$ self-propelled particles (SPPs) in our model we have $\vec{F}_i^{\rm repulsion} = \epsilon \sum_{j \ne i}^N \left(1 - r_{ij}/2r_0\right)^{3/2} \hat{r}_{ij}$, which takes non-zero values only when the distance between two particles $|\vec{r}_i - \vec{r}_j| = |r_{ij} \hat{r}_{ij} | = r_{ij} < 2 r_0$ \cite{Silverberg2013}; $\vec{F}_i^{\rm propulsion} = \mu (v_0 - v_i) \hat{p}_i$, where $v_0$ is a constant preferred speed, $v_i$ is the current speed of the $i^{\rm th}$ SPP, and $\hat{p}_i$ is a unit vector pointing from each particle's center to the common point of interest ${\cal P}$; $\vec{F}_i^{\rm noise} = \vec{\eta}_i$ is a random force vector whose components $\eta_{i,\lambda}$ are drawn from a Gaussian distribution with zero mean and standard deviation $\sigma$ defined by the correlation function $\langle \eta_{i,\lambda}(t) \eta_{i, \kappa}(t') \rangle = 2 \mu \sigma^2 \delta_{\lambda\kappa} \delta(t-t')$, which ensures noise is spatially and temporally decorrelated.   The simulation box's boundaries are rigid so that collisions with SPPs give rise to a force similar to the repulsion force, $\vec{F}_i^{\rm wall} = \epsilon \left(1 - r_{iw}/r_0\right)^{3/2} \hat{r}_{iw}$, which is non-zero when the distance of the particle from the wall $r_{iw} < r_0$, and is directed along the wall's outward normal direction.  In terms of the simulation unit length $\ell$ and unit time $\tau$, we set the particle radius $r_0 = \ell/2$, the simulation box size $L = 50\ell$, the preferred speed $v_0 = \ell / \tau$ \cite{Helbing2000}, the random force standard deviation $\sigma = \ell/\tau^2$ and the force scale coefficients $\epsilon = 25 \ell/\tau^2$, $\mu = \tau^{-1}$ \cite{Silverberg2013}.  Results presented here are for a collection of $N = 200$ SPPs, though varying population size has little effect (Supplemental Materials). 

Simulations were initialized with random initial positions for each particle.  Trajectories were evolved with Newton-Stomer-Verlet integration according to $\ddot{\vec{r}}_i = \vec{F}_i^{\rm repulsion} + \vec{F}_i^{\rm propulsion} + \vec{F}_i^{\rm noise} + \vec{F}_i^{\rm wall}$ for a total of $3,000\tau$ units of time [Fig.~\ref{fig:1}(c)], where each $\tau$ consists of 10 integration time steps.  While data for the initial $\approx 50\tau$ was dominated by transient motion, we discarded the first $300\tau$ from our analysis to avoid far-from-equilibrium effects [Fig.~\ref{fig:1}(c), linear path segments].  By $300\tau$ the SPPs aggregated near ${\cal P}$ and settled into a steady-state configuration with each particle making small random motions about their average positions [Fig.~\ref{fig:1}(d)].  For the model parameters studied here, collisions and random force fluctuations contribute roughly equally to these motions, which can be seen by estimating the relevant time scales.  At average crowd density $n$, the collision time scale is $\tau_{\rm coll} = 1/(2 r_0 v_0 n) \approx (\pi/4) \tau$, the noise time scale is $\tau_{\rm noise} = v_0^2 / 2\mu\sigma^2 = \tau/2$ (Supplemental Materials) \cite{Silverberg2013}, so that $\tau_{\rm coll} \approx \tau_{\rm noise}$ at steady-state.  Thus, while $\vec{F}_i^{\rm propulsion}$ acts as an external field confining SPPs, collision and noise forces are responsible for position fluctuations and the aggregate's disordered structure [Fig.~\ref{fig:1}(d)].

To better understand the role of local structure on global collective motion, we note a striking resemblance between these simulations of high-density crowds and previous studies of disordered packings \cite{Helbing2007,helbing2006analytical,cristiani2011multiscale, faure2015crowd}.  In the context of jammed granular materials, a significant amount of effort has gone into developing theoretical tools that connect local structure to dynamical response \cite{henkes2012extracting, brito2010elementary, ashton2009relationship, manning2011vibrational, henkes2011active, xu2010anharmonic, charbonneau2015}.  A key analysis method involves the displacement correlation matrix whose components are defined by $C_{ij} = \langle \left[\vec{r}_i(t) - \langle \vec{r}_i \rangle \right] \cdot \left[ \vec{r}_j(t) - \langle \vec{r}_j \rangle \right] \rangle$.  Here, $\vec{r}_i(t)$ is the instantaneous position at time step $t$, $\langle \vec{r}_i \rangle$ is the mean position of the $i^{\rm th}$ SPP, and all averages $\langle \cdot \rangle$ were calculated by sampling position data every $10\tau$ for a total of 270 measurements.  This sampling was chosen to reduce effects of auto-correlated motion while still accumulating sufficient statistically independent measurements in a finite time \cite{henkes2012extracting}.  In this computation, we exclude underconstrained SPPs that do not contribute to the overall collective motion.  In the jamming literature these particles are called ``rattlers,'' and they are distinguished by abnormally large position fluctuations \cite{henkes2012extracting}.  In our analysis, we used a position fluctuation threshold of 4 standard deviations to identify rattlers.  However, our results were self-consistent when we varied this parameter from 2 to 5 indicating the methodology is robust to a range of threshold values (Supplemental Materials).

\begin{figure}
\centering
\includegraphics[scale=.98]{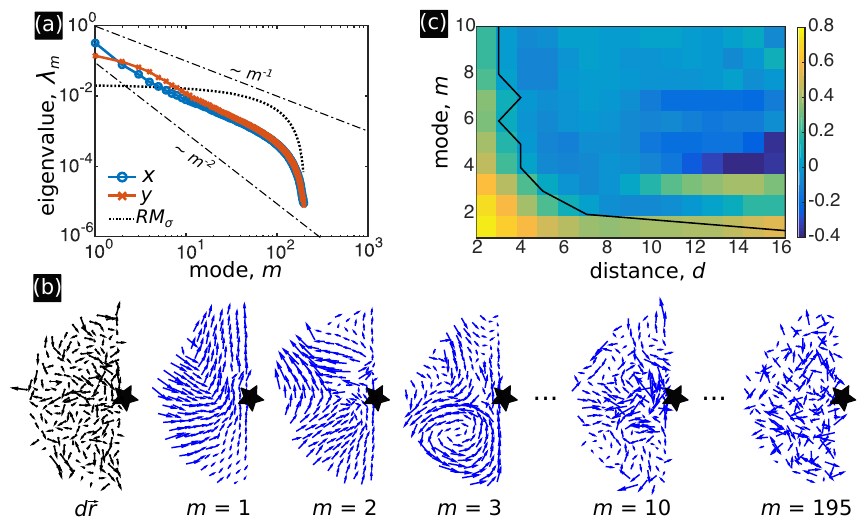}
\caption{\small (color online) Eigenmode analysis of asocial model for high-density human crowds.  (a) Eigenvalue spectrum $\lambda_m$ of the displacement correlation matrix exhibits scaling properties between $\lambda_m \sim m^{-1}$ and $\sim m^{-2}$ (black dashed lines).  Low $m$ eigenmodes in both $x$ (blue) and $y$ (orange) directions are larger than a random matrix model ($RM_{\sigma}$), and thus describe correlated motion. (b) Snapshot of instantaneous displacements $d\vec{r}$ and example vector fields for various eigenmodes.  Lower $m$ eigenmodes are more spatially correlated than higher $m$.  (c) A heatmap of the polarization correlation function for the first 10 eigenmodes as a function of distance $d$ between SPPs.  Black line is where the correlation function decays to 0 demonstrating a long-range highly correlated mode for $m = 1$. }
\label{fig:2}
\end{figure}

To extract quantitative information from the configuration of SPPs, we computed the eigenmodes $\vec{e}_m$ and eigenvalues $\lambda_m$ of the displacement correlation matrix.  In the harmonic theory of crystals, these normal modes fully characterize the linear response of the system to perturbations \cite{ashcroft1976nd}.  For disordered materials, these modes convey information about structural stability as well as coherent  and localized motion \cite{brito2010elementary,ashton2009relationship,manning2011vibrational}.  Plotting the eigenvalue spectrum $\lambda_m$ as a function of mode number $m$ averaged over 10 runs with random initial conditions revealed an approximate power-law decay [Fig.~\ref{fig:2}(a), blue and orange data].  While the Debye model for 2D crystals obeys $\lambda_m \sim m^{-1}$ [Fig.~\ref{fig:2}(a), upper dashed line] \cite{ashcroft1976nd}, the simulation data has an exponent between -1 and -2.  Using a random matrix model of uncorrelated Gaussian variables as a control for decoherent motion [Fig.~\ref{fig:2}(a), black dotted line] (Supplemental Materials) \cite{henkes2012extracting}, we see the lowest six eigenmodes contain information about correlated motion.  Plotting displacement vector fields for a few eigenmodes, we indeed find a higher degree of spatial correlation for lower $m$ that rapidly diminishes with increasing mode number [Fig.~\ref{fig:2}(b)].  To quantify this observation, we measured the polarization of the each mode's vector field and calculated the correlation function for this order parameter (Supplemental Materials) \cite{Cavagna2014}.  Remarkably, we find the first eigenmode carries a system-spanning displacement modulation [Fig.~\ref{fig:2}(c), $m = 1$], whereas the correlation for higher modes rapidly decays over a few particle diameters [Fig.~\ref{fig:2}(c), $m > 1$].  

To understand the origins of this long wavelength mode, we note the point of interest ${\cal P}$ breaks $XY$ translational symmetry, and therefore the Goldstone theorem implies the existence of low-frequency long-wavelength deformations \cite{nambu1960quasi, goldstone1961field, sethna2006statistical}.  This Goldstone mode is expected to arise at low $m$ since eigenvalues are related to vibrational frequencies by $\lambda_m = \omega_m^{-2}$, and the largest eigenvalue in the spectrum occurs at the lowest mode number [Fig.~\ref{fig:2}(a)].  Thus, the system-spanning $m = 1$ eigenmode is the system's Goldstone boson; when excited, it drives the SPPs to move collectively as one \cite{ramaswamy2010mechanics}.  An example of a disaster resulting from this type of coherent long-range motion is known as ``crowd crush'' \cite{fruin1993causes}.  In these situations, a large number of people are suddenly displaced toward a wall, fence, or other architectural element resulting in dangerously high pressures \cite{Helbing2007}.  As a consequence, injuries and death are known to occur.  Determining if Goldstone modes are responsible for crowd crush would require careful image analysis of crowd structure and motion in the moments before such an event.  Nevertheless, we expect any large dense gathering of people to exhibit this type of long-range collective behavior since its origins can be traced to the general principle of symmetry breaking.

\begin{figure}
\centering
\includegraphics[scale=.98]{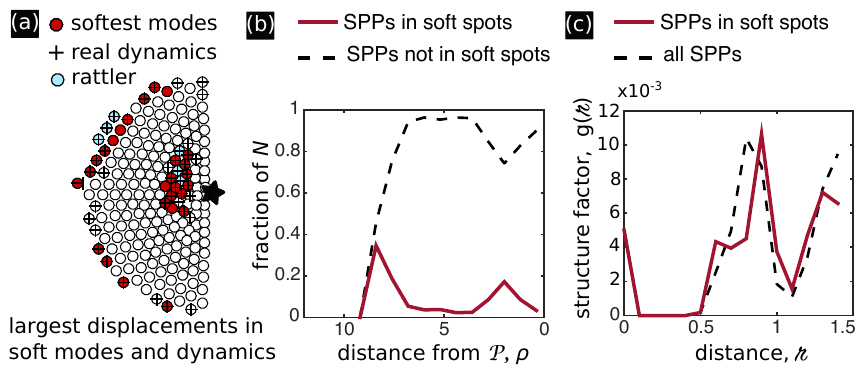}
\caption{\small (color online) Soft spots within the crowd undergo large displacements.  (a) SPPs are shown as disks.  Soft spots near the core of the aggregate colocalize with SPPs that displace the most in real dynamics.  This region is subject to large structural rearrangements when the system is perturbed, and is likely a region where injury can occur.  Apparent soft spots along the periphery are artifacts due to underconstrained edge effects.  (b) Averaging over all simulation runs show soft spots generally occur near the core of the aggregate a radial distance $\rho \approx (2 \pm 1)$ away from ${\cal P}$.  (c) Structure factor $g(\scripty{r})$ measures the pair-wise SPP distribution and reveals structural features distinguishing SPPs in soft spots that suggest why they are subject to large displacements. }
\label{fig:3}
\end{figure}

Another type of disaster found at high-density social gatherings is when sudden unexpected movements of the crowd cause individuals to trip and fall.  Because the majority of people are unaware this accident has happened, the rest of the crowd continues to move largely uninterrupted, resulting in injury or death due to trampling or compressive asphyxia \cite{fruin1993causes, Helbing2007, krausz2012loveparade}.  This is more general than the excitation of a pure Goldstone mode, and is better characterized by a superposition of modes.  Thus, we focus on the particles that displace significantly more than average in a given mode $m$ [Fig.~\ref{fig:3}(a), displacement threshold is 2.5 standard deviations more than average] (Supplemental Materials).  Studies of jammed granular media show these particles, which tend to cluster in regions called ``soft spots,'' correlate with structural rearrangements when the system is perturbed \cite{manning2011vibrational, charbonneau2015}.  Superimposing data from the first 10 modes of a single simulation run reveals a soft spot near the core of the aggregate [Fig.~\ref{fig:3}(a)].  Regions along the perimeter also featured large displacements, but they are essentially underconstrained edge effects and therefore not relevant for our analysis.  Identifying SPPs undergoing the largest displacements in each mode up to $m = 10$ in all simulation runs showed the region near the core of the crowd is the most likely area to find soft spots [Fig.~\ref{fig:3}(b), peak centered on $\rho \approx 2$].  Cross-correlating soft spot SPPs with their real-space dynamics confirmed these particles typically displace the greatest amount despite being confined within a disordered aggregate [Fig.~\ref{fig:3}(a)].

We further studied the relation between structural disorder and large displacements in soft spots by measuring the pairwise distribution $g(\scripty{r})$ as a function of distance $\scripty{r}$ between particles (Supplemental Materials) and found that soft spot SPPs have an intrinsically different structure compared to the average population [Fig.~\ref{fig:3}(c)].  The plateaued region in $g(\scripty{r})$ around $0.5 \lesssim \scripty{r} \lesssim 0.8$ [Fig.~\ref{fig:3}(c) full line] indicates soft spot SPPs are more highly squeezed by some of their neighbors, while the shifted peak centered on $\scripty{r}~\approx~0.9$ indicates they're also further away than average from other neighbors [Fig.~\ref{fig:3}(c), dashed line peak at $\scripty{r} \approx 0.8$].  These data suggest soft spot SPPs are being compressed tightly in one direction, and as a consequence displace greater amounts in the orthogonal direction.  As such, structural disorder is fundamental for  large displacements and rearrangements [Fig.~\ref{fig:3}(a)] \cite{manning2011vibrational}.  Thus, we hypothesize soft spots in human crowds pose the greatest risk for tripping and subsequent trampling.  If found true, real-time image analysis identifying soft spots in densely-packed human crowds may provide useful predictive power for preventing injuries.

\begin{figure}
\centering
\includegraphics[scale=.98]{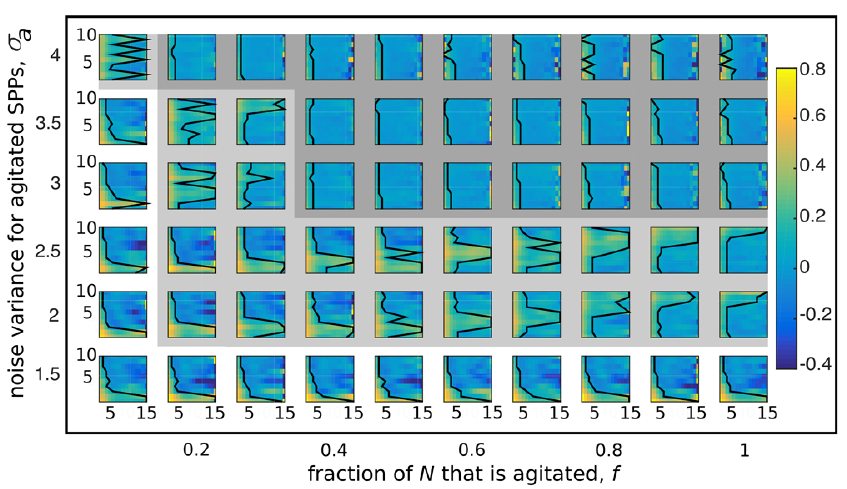}
\caption{\small (color online) Introducing a fraction $f$ of agitated SPPs with variance $\sigma_a$ in $\vec{F}_i^{\rm noise}$ to the total population $N$ probes structural origins of collective motion.  Each heat map is the polarization correlation function for the first 10 eigenmodes as a function of distance $d$ (same as Fig.~\ref{fig:2}(c)). Low fluctuations (white background) preserve the long-range highly-correlated Goldstone mode near $m = 1$.  High fluctuations (dark gray background) destroy long-range correlated modes.  Intermediate fluctuations (light gray background) add new modes with long-range correlations, indicating stochastic resonance. }
\label{fig:4}
\end{figure}

Our results thus far have focused on structural origins of collective motion with all model parameters kept constant.  In real life situations, not all people behave the same: some agitate more easily, others less so \cite{heide2004common, Helbing2000PRL}.  Accordingly, we modify the asocial model to study how mechanisms for coherent collective motion are affected by active perturbations from within.  Specifically, we introduce a second population of SPPs so that a fraction $f$ exhibits a more agitated behavior, while the remaining fraction $1-f$ of the population are the same as before \cite{Silverberg2013, Helbing2000PRL}.  We model these agitated SPPs with a larger distribution of force fluctuations in $\vec{F}_i^{\rm noise}$ by increasing their standard deviation to $\sigma_a > \sigma$, and analyzing the two parameter phase space made of $f$ and $\sigma_a$.  We first consider the case $\sigma_a = 3\sigma$ and vary $f$ from 0 to 1.  Calculating the spectrum of eigenvalues $\lambda_m$ shows the qualitative trends are independent of $f$, though numerical values of $\lambda_m$ tend to increase with more agitators (Supplemental Materials).  To understand how long-range collective motion is affected by agitated SPPs, we measured the polarization correlation function for the first 10 modes by varying $\sigma_a$ and $f$ [Fig.~\ref{fig:4}].  Surprisingly, the correlation functions for $\sigma_a = 3\sigma$ at various values of $f$ show a qualitative transition unanticipated from the eigenvalue spectrum.  For $f = 0.1$, a long-range correlated Goldstone mode is observed as before.  However, multiple long-range correlated modes are observed for $f = 0.2$, and no long-range correlated modes are observed for $f > 0.3$.  Examining other values of $\sigma_a$ shows a similar transition with increasing $f$ from a single well-defined long-range mode, to multiple long-range modes, to no long-range modes whatsoever [Fig.~\ref{fig:4}, rows left-to-right].

The low-agitation and high-agitation limits are intuitive.  For low agitation [Fig.~\ref{fig:4}, white region], additional force fluctuations through increasing $\sigma_a$ with low $f$ or increasing $f$ with low $\sigma_a$ induce small perturbations to the overall structure.  As such, the existence of a Goldstone mode at low $m$ is anticipated based on the homogeneous population results [Fig.~\ref{fig:2}(c)].  For high agitation where the combined effect of $\sigma_a$ and $f$ is large [Fig.~\ref{fig:4}, dark gray shaded region], we expect local structure of the aggregated SPPs to break down and correlated motion to be marginalized.  Consistent with this reasoning, we find no long-range modes in the high-agitation limit (Supplemental Materials).  

Between the high and low agitation limit, we find a boundary in the $(f, \sigma_a)$ phase diagram characterized by multiple long-range modes [Fig.~\ref{fig:4}, light gray shaded region].  This result is striking because it shows moderate levels of noise induces new coherent modes.  Noting that correlated motion allows mechanical information to be transferred across the aggregate, an appearance of multiple long-range modes implies greater information bandwidth.  In certain settings, signal enhancement mediated by noise is called \textit{stochastic resonance} \cite{mcdonnell2009stochastic, gammaitoni1998stochastic}.  Stochastic resonance can be found in systems where nonlinear effects dampen signal propagation, but by introducing random noise, the effects of nonlinear terms are reduced leading to a restoration of signal propagation.  In our case, nonlinear effects come from structural packing disorder that suppresses conventional phonon modes found in ordered 2D systems.  Random noise from agitators increases an internal pressure \cite{Takatori2014} within the aggregate that helps break-up this heterogeneous structure.  Consequently, additional phonon modes are able to reassert their presence.  In the context of our model, this finding means that modest random fluctuations can enhance overall collective motion, which increases the potential for injurious outcomes in high-density crowds.  

Our analysis of collective motion in dense crowd simulations relies on trajectory data in order to identify and understand the emergence of Goldstone modes, soft spots, and stochastic resonance.  With an eye to crowd safety, the dependence on readily measurable quantities combined with computer vision techniques \cite{junior2010crowd, ali2013modeling} provides significant potential for applications in real-time crowd management.  In the long-run this may help protect attendees at large gatherings by reducing emergent risks \cite{Helbing2007, krausz2012loveparade, heide2004common}.  More theoretically, the observation of Goldstone modes hints that a collective motion analogous to the Higgs particle may also be found in future studies of crowd speed modulations.  Indeed, developing an effective field theory with quasi-particle-like excitations could present new opportunities to understand emergent collective motions, their interactions, and potential hazards.

\section*{Acknowledgements}
We thank U. Biets and J. Bailey for providing photographs used here. A.B. thanks A. Maffini for discussions during the early stages of the project. J.L.S. thanks M. Bierbaum, J. Sethna, and I. Cohen for useful discussions. Thanks to A. G\aa din and J. Svensson for software development.  A.B. acknowledges funding from the Centre for Interdisciplinary Mathematics (CIM). J.L.S. was independently funded.

\bibliographystyle{apsrev4-1}
%

\pagebreak
\begin{widetext}
\begin{center}
\textbf{\large Supplemental Materials: Emergent Structural Mechanisms for High-Density Collective Motion Inspired by Human Crowds}
\end{center}
\end{widetext}
\setcounter{equation}{0}
\setcounter{figure}{0}
\setcounter{table}{0}
\setcounter{page}{1}
\makeatletter
\renewcommand{\theequation}{S\arabic{equation}}
\renewcommand{\thefigure}{S\arabic{figure}}
\renewcommand{\bibnumfmt}[1]{[S#1]}
\renewcommand{\citenumfont}[1]{S#1}

\section*{Methods}

\paragraph{Simulations.} 
Each simulation takes place in a square room of side $L=50$ length units $\ell$, the center of the room is placed in the origin [0,0]. $N$ individuals are modeled as self-propelled particles (SPPs) of radius $r_0 = 0.5\ell$ are placed in random initial positions and with 0 initial speed. 
The equation of motion is numerically integrated using the Newton-Stomer-Verlet algorithm.
At each time step the full algorithm computes the total force $\vec{F}_{\text{tot}}(t)$ acting on each individual in its current position $\vec{r}(t) $. The position is updated using the current speed $\vec{v}(t) $ according to
\begin{equation}
\vec{r}(t+1) = \vec{r}(t) + \vec{v}(t)   \Delta T + \frac{1}{2} \vec{F}_{\text{tot}}(t) (\Delta T)^2. 
\end{equation}
The algorithm then computes the force $\vec{F}_{\text{tot}}(t+1)$ acting on each individual in its new position, and updates the speed:
\begin{equation}
\vec{v}(t+1) = \vec{v}(t) + \frac{1}{2} (\vec{F}_{\text{tot}}(t) + \vec{F}_{\text{tot}}(t+1)) \Delta T. 
\end{equation}
We choose $\Delta T = 0.1$. Such a value is small enough to make the trajectories smooth, but large enough to achieve a reasonable computational time. 
Every 10 time steps of the simulation we record each particle's position and the pressure experienced due to radial contact forces $P_i = (2 \pi r_0)^{-1}  ( \sum_j F_{i,j}^{\text{repulsion}} +  \sum_w F_{i,w}^{\text{wall}})$. We run our simulations for $T = 30,000$ time steps, corresponding to $3,000 \tau$.

\paragraph{Model time scales.}  For the model parameters studied here, collisions and random force fluctuations contribute roughly equally to these motions, which can be seen by estimating the relevant time scales.  In this case, the collision time scale $\tau_{\rm coll} = 1/(2 r_0 v_0 n) \approx (\pi/4) \tau$ is the mean-free path $(2 r_0 n)^{-1} \approx (\pi/2) r_0$ divided by the preferred speed $v_0$, where an estimate of the average crowd density $n \approx N / \pi (\sqrt{N} r_0)^2$ is obtained by noting the steady-state configuration of SPPs is roughly a half-circle with radius $\sqrt{N}r_0$ surrounding ${\cal P}$.  Similarly, the noise time scale $\tau_{\rm noise} = v_0^2 / 2\mu\sigma^2 = \tau/2$ can be found calculating the amount of time required for noise to change the correlation function $\langle [ v_i(\tau_{\rm noise}) - v_i(0)]^2 \rangle = 2 \mu \sigma^2 \tau_{\rm noise}$ by an amount equal to the characteristic speed squared.  Consequently, $\tau_{\rm coll} \approx \tau_{\rm noise}$ at steady-state.

\paragraph{The correlation matrix.}  As the starting point of our analysis, we use the simulated trajectories to compute the displacement covariance matrix $C_p$ \cite{henkes2012extractingSM}. 
We treat separately the $x$ and $y$ components of the position vector $\vec{r}_i(t) = [ x_i(t),y_i(t) ]$.
The simulation reaches steady state after $\approx 50 \tau$, but we discard the first $300 \tau$ to eliminate far-from-equilibrium transients. 
Of the remaining $2,700\tau$, we sample data every $10 \tau$, so that we have $270$ time points. The obtained time points are separated by 100 simulation time steps to ensure statistical independence \cite{henkes2012extractingSM}. 

The equilibrium position $\langle x_i \rangle$ is obtained by averaging each SPPs position over $270$ time points. The displacements $\delta x_i(t) = x_i (t) - \langle x_i \rangle$ around the mean position are used to compute the correlation matrix at the sampled time steps. 
The covariance matrix for the simulation is obtained by averaging over the independent time points:
\begin{equation}
C_{p_x} = \langle [x_i (t) - \langle x_i \rangle ] \cdot [x_j (t) - \langle x_j \rangle ] \rangle,
\end{equation}
with a similar computation for the $y$ component. The 270 independent time points are sufficient for the covariance matrix to converge to the true correlation matrix of the underlying statistical process \cite{henkes2012extractingSM}. In other words, the averaging is sufficient to obtain equivalence between time and ensemble average.  From now on we shall refer to $C_p$ as correlation matrix.

\paragraph{Interpretation of $C_p$.} For thermally equilibrated systems and in the approximation of harmonic oscillations, the dynamical matrix $D$  (the Hessian of the pair interaction potential divided by the mass) contains all the information about the time-evolution of the system. 
The eigenmodes of $D$ represent the vibrational modes of the system, $D \vert \lambda_m \rangle = \omega_m^2 \vert \lambda_m \rangle$, where $\omega_m$ are the vibrational frequencies of the system, and $\omega_m^2$ can be interpreted as the energy that has to be transferred to the system to activate the corresponding vibration.

For these systems, the correlation matrix is proportional to the inverse of the dynamical matrix $D$, thus the eigenvectors of $C_p$ are simultaneous eigenvectors of $D$, while their eigenvalues are inversely proportional and $\omega_m^2 \sim 1/\lambda_m$. 
For thermal systems at equilibrium, the spectrum of $C_p$ can then be interpreted as the vibrational modes of an equivalent system of harmonic springs (shadow system). 
The vibrational properties of the two systems might not be exactly the same, but studying the shadow system is enough to extract the properties of the real system.

In our case, we are dealing with active matter and the observed dynamics is the result of the interplay of propulsion, repulsion, noise forces, and the environmental constraint of walls.
Our system is not at thermal equilibrium, and the spectrum of $C_p$ cannot be strictly interpreted as vibrational modes.
We thus perform a Principal Component Analysis (PCA) of the covariance matrix to extract the components of the fluctuations that carry the information of correlated motion by comparison with the random matrix case $RM_{\sigma}$.

\begin{figure}
\centering
\includegraphics{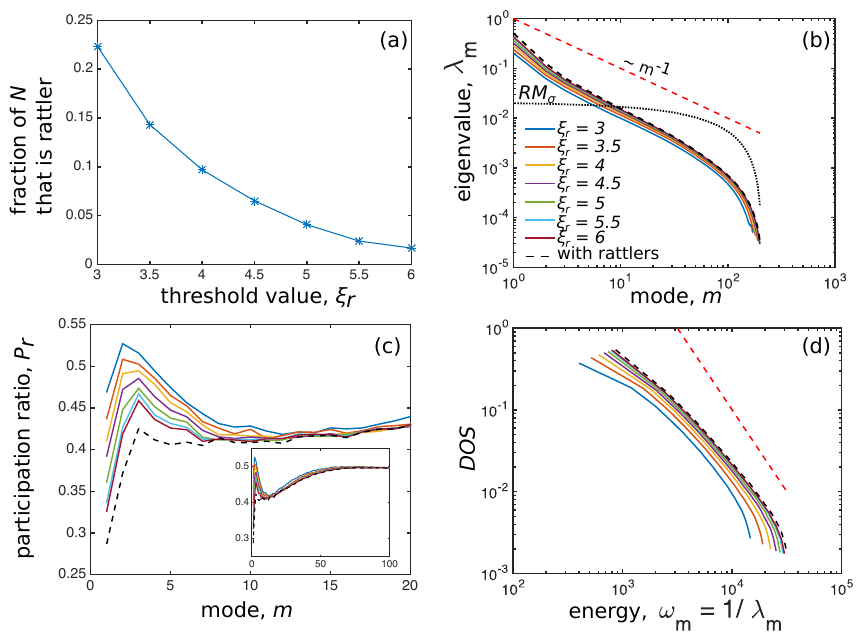}
\caption{\small Outcome of the choice of different thresholds when looking for rattlers and comparison with the random matrix model ($RM_{\sigma}$, dotted line). (a) The fraction of rattlers is a monotonically decreasing function of the threshold value $\xi_r$.  (b) Comparison of the spectrum of the eigenvalues with the random matrix case before removing the rattlers (dashed) and after at different values of the threshold. At increasing $\xi_r$ the first five to ten modes are larger then the random matrix case.
 (c) The Participation ratio increases for the lowest energy modes when rattlers are removed at all values of the threshold $\xi_r$.  (d) Comparison of the density of states before removing rattlers (black dashed line) and after (solid colored lines) at different values of the threshold $\xi_r$.  Debye model for harmonic crystals (red dashed line) is shown for reference.}
\label{fig:rat_thr}
\end{figure}

\paragraph{The Random Model $RM_{\sigma}$.}
We use a Random Model of uncorrelated Gaussian
variables to test what are the relevant eigenmodes in our system. We compute the correlation matrix of a set of random displacements normally distributed with zero mean and variance $\sigma_{RM}$. The variance is the same as the simulated displacements around each SPPs equilibrium position.
The eigenmodes computed from $C_p$ with eigenvalues larger than the largest one of the $RM_{\sigma}$ model contain the relevant information about correlations. In our case we see that the first six modes are above noise (before removing the rattlers, Fig.~\ref{fig:rat_thr}).

\paragraph{Measures.} For each component of the correlation matrix we compute the eigenvalues $\lambda_m$ and the eigenvectors $\vec{e}_m^i$, $i, m =1 \dots N$ in the $x$ and $y$ directions for a total of $2N$ eigenvalues.
The eigenvalues taken in decreasing order and plotted as a function of their index $m$ give the spectrum of the correlation matrix (for separated components $x$ and $y$).
Using the analogy with vibrational theory, we call $\omega_m^2 = \lambda_m^{-1}$ the energy of the $m$-th mode, from which one could compute the density of states, DOS$(\omega^2)$.
The DOS carries information about the rigidity of a solid to collective motion, but since the equipartition of energy is violated for active matter, one should be careful when analyzing this quantity.  The participation ratio $P_r(m) = (\sum_{i=1}^{N}  \vert e_m^i \vert)^2/[ N (\sum_{i=1}^{N}  \vert e_m^i \vert^4)]$, with $ \vert e_m^i \vert = ((e_{m_x}^i)^2 + (e_{m_y}^i)^2)^{1/2}$\cite{xu2010anharmonicSM} is constructed by combining the $x$ and $y$ components.  In crystal theory the participation ratio of a mode describes how many particles in the system move in a given mode, and runs between 0 (fully localized) to 1 (fully extended).
If we think about modes as collective dynamics, another useful characterization of their collective nature and spatial coherence is given by the mean polarization $\vec{\Phi} (m) = N^{-1} \sum_{i=1}^N \vec{e}_m^i/ \vert \vec{e}_m^i \vert $ and the correlation function of the fluctuations around it
\begin{equation}
C_m(d) = \langle (\vec{e}_m^i - \vec{\Phi}) (\vec{e}_m^j - \vec{\Phi}(m)) \rangle_{d_{i,j}=d},
\end{equation}
From this computation, we define the correlation length $l_{c}(m)$ such that $C_m(l_{c})=0$ \cite{Cavagna2014SM}.  In order to characterize the structure around SPPs in soft spots, we use the two particle radial structure factor $g(\scripty{r}) = \sum_{i=1}^{N} \sum_{j \neq i}^{N} \delta(\scripty{r} - r_{ij}) / [N (N-1)]$, which measures the radial distribution of neighboring SPPs.  In our analysis, the eigenvalue spectrum, density of states, participation ratio, correlation function, and structure factor are averaged over 10 independent simulations of the dynamics with random initial conditions.

\paragraph{Choice of the rattler's threshold.}
In the literature, rattlers are underconstrained particles that feature an abnormally large displacement in the lowest energy modes \cite{henkes2012extractingSM}. They are a consequence of local structure: neighbors pack inhomogeneously and form cages where these particles get trapped and vibrate without participating in collective movement.  Following the analysis developed for granular materials \cite{henkes2012extractingSM}, we identify rattlers from the eigenmode analysis, eliminate these particles from our consideration, and recompute the eigenmodes and eigenvalues for a new $C_p$ that only considers the population of non-rattlers. This procedure is necessary so that the lowest energy modes are not zero energy vibrations localized on a single particle (i.e. the rattler).  Instead, this two-step computation of $C_p$ ensures we accurately measure collective motion of the system.

The identification criteria for rattlers we utilize comes from previous work \cite{henkes2012extractingSM}: a particle $i$ is a rattler if $dr_m^i \geq \langle dr_m \rangle + \xi_r \sigma_m$, where $dr_m^i =  \vert e_m^i \vert = ((e_{m_x}^i)^2 + (e_{m_y}^i)^2)^{1/2}$ is the displacement of the particle $i$ on the $m$-th mode, $\langle dr_m \rangle$ is the average displacement of the particles on that mode, $\sigma_m$ their standard deviation, and $\xi_r$ is a fixed threshold.
In order to fix a value for $\xi_r$ we compute $C_p$, its eigenvalues and eigenvectors.
We identify rattlers in the eigenmodes corresponding to the first 10 modes with $\xi$ from 2 to 5. 
We consider the first ten modes as they are the ones out of random noise.  After identifying the rattlers, we eliminate them and re-compute $C_p$, its eigenvalues and eigenvectors. 
We compute the spectrum, the DOS and the participation ratio and we average over 10 independent instances.

We compare the above measures with and without the rattlers for different values of the threshold [Fig.~\ref{fig:rat_thr}].
The trend of the spectrum of the eigenvalues and of the density of states is not affected by removing the rattlers. In particular, we find the first modes remain above random noise [Fig.~\ref{fig:rat_thr}(b)]. The participation ratio [Fig.~\ref{fig:rat_thr}(c)] is increased at all values of $\xi_r$, suggesting that the rattlers we removed were particles that fluctuated the most in the considered modes.  Thus, we fix the rattlers threshold at $\xi_r = 4$. With this value the rattlers are about the $5\%$ of the total population [Fig.~\ref{fig:rat_thr}(a)].

\paragraph{Choice of threshold for soft spots.}

\begin{figure}
\centering
\includegraphics[scale=.98]{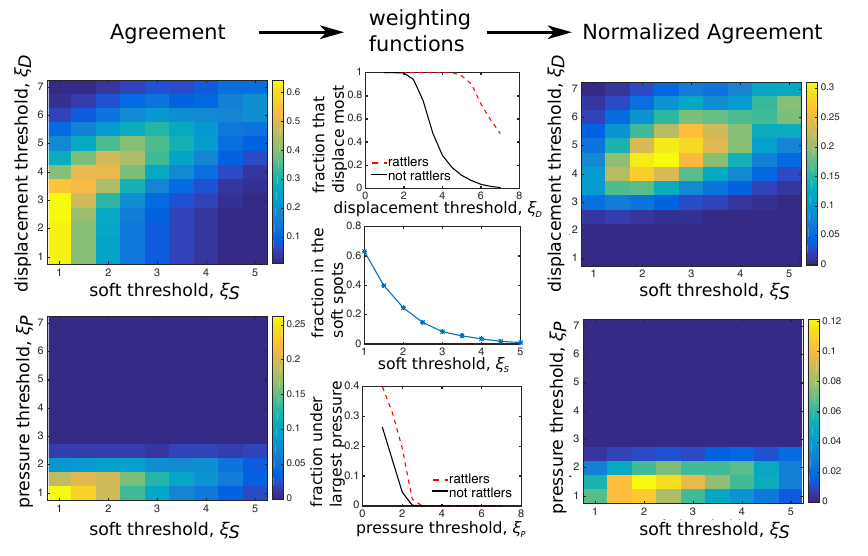}
\caption{The agreement functions before (left column) and after (right column) the normalization through the weighting functions (central column) that measure the size of the sets $S, P, D$.}
\label{fig:soft_thr}
\end{figure}

\begin{figure}
\centering
\includegraphics[scale=.98]{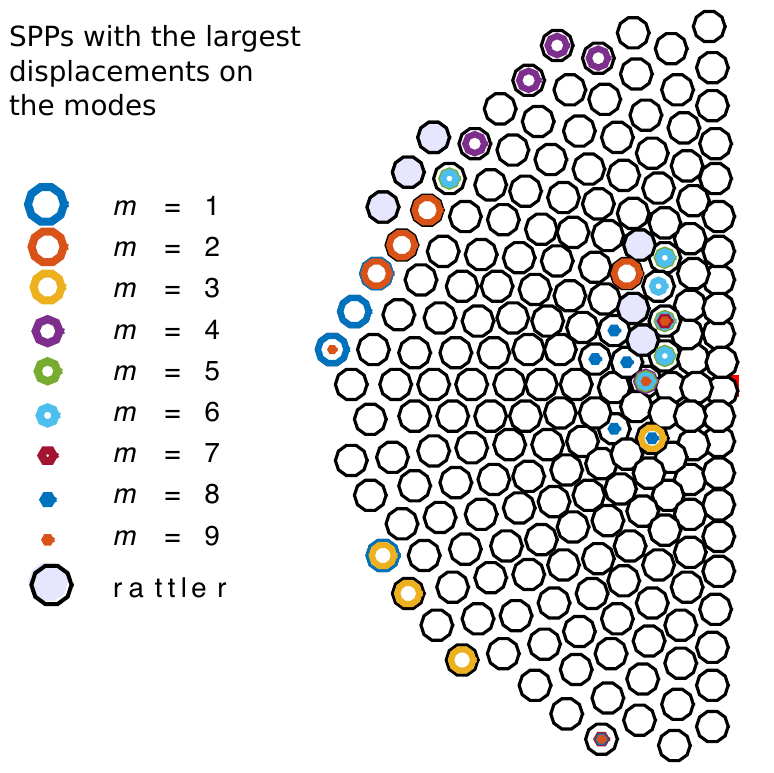}
\caption{Each SPP is shown as a disk. Superimposing data from this simulation run for the first 10 modes shows that soft spots arise in multiple modes overlapped in the same area. Apparent soft spots along the periphery are primarily artifacts due to underconstrained edge effects.}
\label{fig:soft_spots}
\end{figure}

As for rattlers, we need to set a threshold for identifying particles featuring large displacement on the softest modes. 
These particles are the ones that displace most when a mode is activated and we might expect them to behave qualitatively differently than the other particles. 
In particular we test if the particles participating the most in soft modes ($S$) are the same particles featuring a large real displacement ($D$) and, separately, if they are subject to large pressure ($P$).
We use the same criteria as for identifying rattlers and test a set of thresholds $\xi_*; \, * = S, D, P $ from 1 to 5 standard deviations with respect to the mean value. 
Once identified the sets corresponding to each threshold, we compute their ``normalized agreement'', as $\nu_S \nu_D \vert S \cap D \vert / \vert S \cup D \vert$. 
$\nu_* = 1 - N_* / N$ is a weighting function that dampens the measure of the overlap if the two sets are oversampling the total population. 

For a large set of thresholds for $D$ and $S$, the particles that participate most in the soft modes are also the ones that feature the largest displacements in the real trajectory.
The normalized agreement function for $S$ and $D$ is maximum for $\xi_S= 2.5$ and $\xi_D = 4.5$ [Fig.~\ref{fig:soft_thr}, top] 
Thus, we identify soft spot SPPs as the ones that have displacement larger than 2.5 standard deviations on at least one soft mode.
Experiencing large pressure ($P$) does not seem to correlate with being a soft spot SPP [Fig.~\ref{fig:soft_thr}, bottom]. 

\section*{Size effect}

\begin{figure}[t!]
\centering
\includegraphics{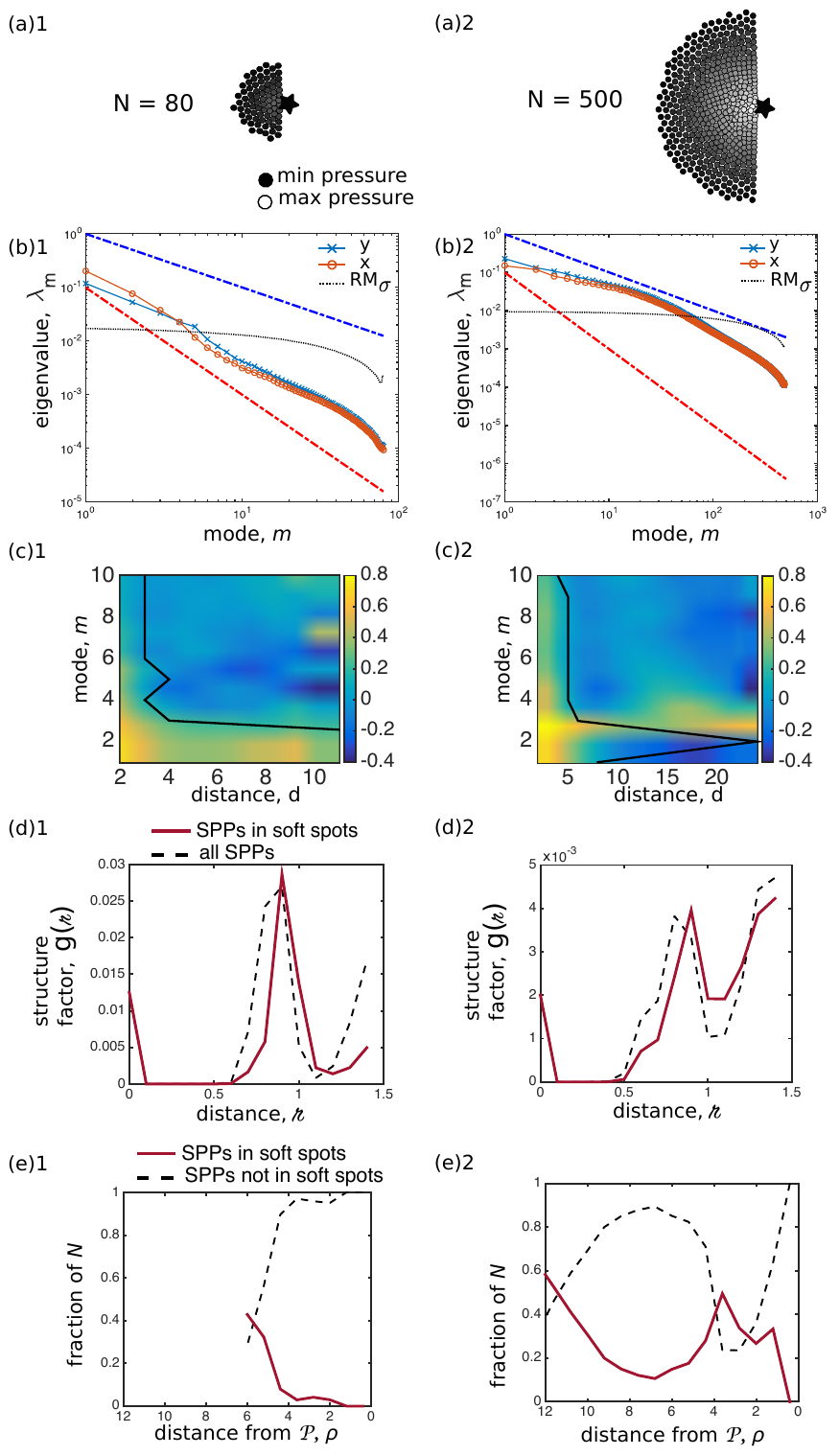}
\caption{\small Main measures at small ($N = 80$, column 1) and large ($N = 500$, column 2) system size.
(a). Each circle represents a SPP, the colorscale is normalized in both (a)1 and (a)2 so that the maximum pressure (lightest circles) is $56.5 P_0$ and the minimum pressure is 0;
(b). Spectrum in the $x$ (circles) and $y$ (crosses) direction compared with the random matrix model (dotted line);
(c). Correlation function (heatmap) and correlation length (line). The gradient scale is the same for (c)1 and (c)2;
(d). Structure factor for the particles in the central soft spot ](full line) compared with the average structure (dashed line);
(e). Proportion of particles in the soft spots (full line) and outside the soft spots (dashed line) as a function of the distance from ${\cal P}$ ($\rho$); 
}
\label{fig:sizefx}
\end{figure}

To check for finite size effects in our analysis we simulated the same dynamics as described in the main text for $N = 500$ and $N = 80$. 
We apply the same procedure as in the case $N = 200$ and we consider the measures that are relevant for the analysis described in the main text [Fig.~\ref{fig:sizefx}].
Pressure increases with size from $23.8 P_0$ for $N = 80$ to $56.5 P_0$ for $N = 500$, where $P_0 = v_0 (2 \pi r_0)^{-1}$ is the inertial pressure of a SPP of radius $r_0$ and moving with speed $v_0$. Our result is in agreement with the empirical observation that crowd pressure builds up with the number of people involved in pushing. 
The eigenvalue spectrum preserves its shape and thus seems to be a genuine feature of the dynamics. For increasing $N$, additional modes rise above the random matrix model for uncorrelated noise, $RM_{\sigma}$.  The lowest energy modes are still correlated over long ranges. In particular for $N = 500$ the first mode shows high correlation at short distance and high anti-correlation at long distance, while the second mode's correlation length spans the size of the system. However the energy gap between these two modes is smaller than for smaller system sizes. For both $N = 500$ and $N = 80$ the structure factor $g(\scripty{r})$ shows that the particles belonging to the core soft spot are slightly less constrained than the average structure.  At the edge of the simulated crowd, the number of SPPs with large displacement on the softest modes is large in both cases, while the size of the core soft spot increases with $N$.

\section*{Correlation with increased activity}

\begin{figure*}
\centering
\includegraphics[width=1\textwidth]{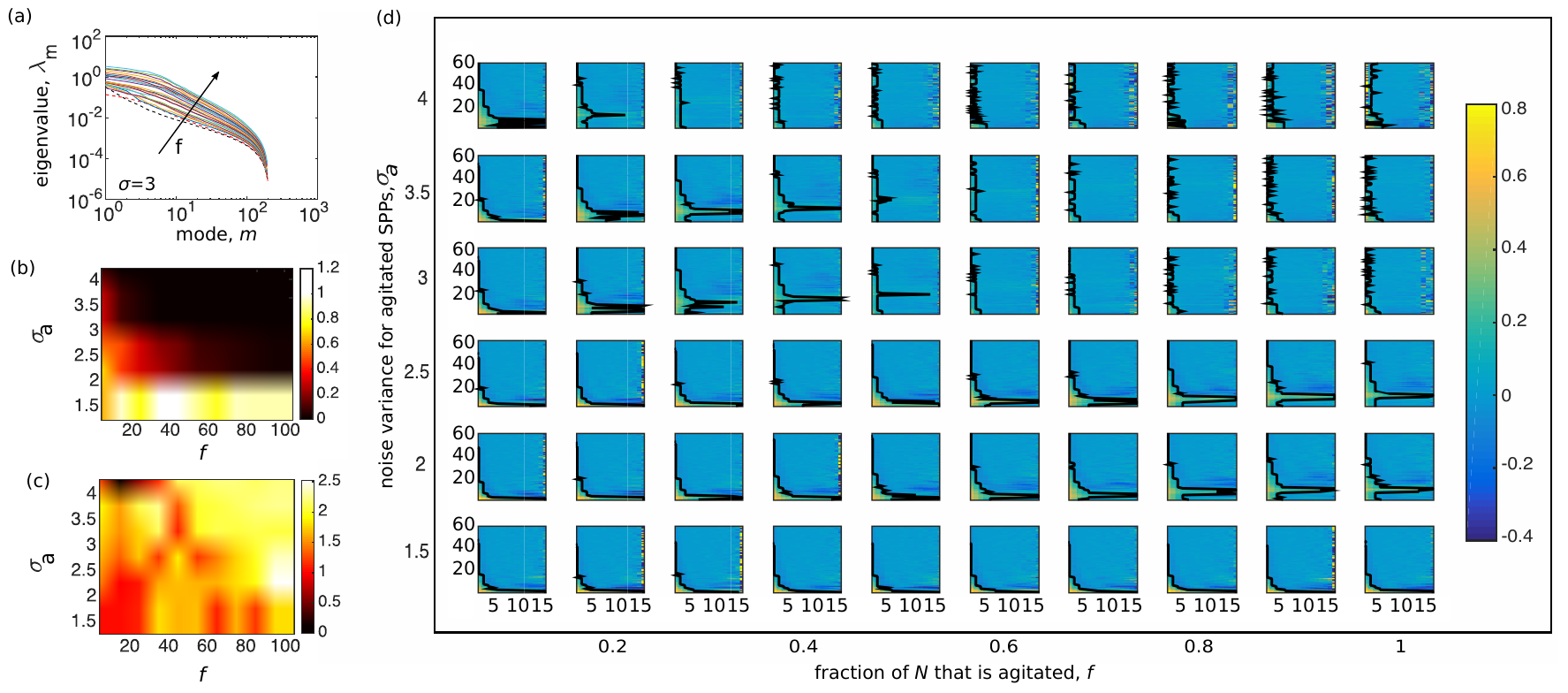}
\caption{\small Effects of increasing the fraction of active people $f$ featuring a larger noise strength $\sigma_a$.
(a) At fixed $\sigma_a = 3$, increasing $f$ increases all eigenvalues of the spectrum with respect to the homogeneous population where $f = 0$. 
(b) The sum of the excitation energies of the first five eigenmodes as a function of $f$ and $\sigma_a$ divided by the corresponding quantity at $f = 0$. Excitation energies significantly drop at increasing active population and activity level.
(c) The size of soft spots at different $f$ and $\sigma_a$ divided by the average soft spot size in the homogeneous population is an increasing function of $f$ and $\sigma_a$.
(d) Correlation function of the fluctuations around the mean polarization order parameter with increasing $f$ and $\sigma_a$ up to the $60^{\rm th}$ mode (heatmap).  The correlation length (black line) is plotted as a function of mode number $m$ in each subplot.  }
\label{fig:longcorr}
\end{figure*}

We model behavioral diversity by increasing noise fluctuations $\sigma_a \in [1.5, 4]$ for different fractions of SPPs $f \in [0.1, 1]$ placed in random positions within the crowd.
When looking at the spectrum of $C_p$, the eigenvalues steadily increase at increasing activity level $\sigma_a$ and at increasing active fraction $f$ [Fig.~\ref{fig:longcorr}(a)], with more and more modes becoming relevant in describing the dynamics.  Correspondingly, the energy needed to excite these modes drops significantly. 
For example, the five lowest energy modes at $f = 0.1$ and $\sigma_a = 2.5$ need half of the excitation energy that is needed when everyone is calm [Fig.~\ref{fig:longcorr}(b)]. 
This means that a fixed amount of energy provided to the system would excite an increasing number of motion modes as more active people are present.
This could be explained by the increase in the size of soft spots, i.e. the number of individuals composing them, with both $f$ and $\sigma_a$ [Fig.~\ref{fig:longcorr}(c)]. 
Intuitively, one might expect that individuals with larger $\sigma_a$ would be detected as rattlers or as belonging to a soft spot, having a larger displacement on the softest modes.
This does not seem to be the case as unstable areas can double their size even when a small fraction of the population becomes very active, as for example for $f=0.1$ and $\sigma_a=3.5$.

The heatmap representing the correlation function for the fluctuations around the mean polarization for eigenmodes up to $m=60$ [Fig.~\ref{fig:longcorr}(d)] expands on the corresponding figure in the main text and confirms our interpretation for modes with $m > 10$.  At increasing $f$ and $\sigma_a$ we observe that coherence decreases for low energy modes and increases for higher energy modes. 
Interestingly, for intermediate values of $\sigma_a$ and $f$ several modes feature long range correlation.
When both $\sigma_a$ and $f$ are very large ($\sigma_a=3,3.5$, $f>0.6$ and $\sigma_a=4$, $f>0.1$) our analysis does not find long-range correlated modes.

\section*{Materials}
Crowd simulations are performed in C. 
The program has been adapted from the original version by Andreas G\aa din and John Svensson.  Data analysis was performed with MATLAB.

\bibliographystyle{apsrev4-1}

\end{document}